# 3D Printing via material extrusion on an acoustic air bed

Sam Keller[1], Matthew Stein[1], Ognjen Ilic[1*]

[1]Department of Mechanical Engineering, University of Minnesota, Minneapolis, 55455, USA

**Abstract:** Additive manufacturing, such as 3D printing, offers unparalleled opportunities for rapid prototyping of complex three-dimensional objects, but typically requires simultaneous building of solid supports to minimize deformation and ensure contact with the printing surface. Here, we theoretically and experimentally investigate the concept of material extrusion on an "air bed" – a judiciously engineered acoustic field that supports the material by contactless radiation force. We study the dynamics of polylactic acid filament (PLA)—a commonly used material in 3D printing—as it interacts with the acoustic potential during extrusion. We develop numerical models to determine optimal transducer arrangements and printing conditions, and we build and demonstrate a concept prototype that integrates a commercial 3D printer and open-source control code. Our results point towards alternative, contactless support mechanisms with potential benefits such as fewer surface defects, less material waste, lower cost, and reduced manufacturing time. These features could become crucial as additive manufacturing continues to evolve into a foundational tool in engineering and beyond.

## INTRODUCTION

In additive manufacturing, the creation of complex objects relies on employing effective support structures during the manufacturing process. In processes such as filament extrusion and material jetting, solid supports are often printed simultaneously with the desired object to ensure stability and proper adhesion to a horizontal printer bed. These supports, along with a solid printing bed, can cause significant defects on the surface of objects, lead to warping and deformation, and prevent their detachment from the bed. Additionally, support structures generate considerable material waste, adding to overall manufacturing time and costs. As additive manufacturing technology rapidly evolves, developing more efficient support mechanisms could have a critical impact on 3D printing. These alternatives must account for the complex nature of the filament as it is melted and extruded to ensure minimal deformation as the filament hardens. Here, we study acoustic levitation as a viable mechanism for contactless filament support during extrusion. By directing a tailored acoustic wavefront at the filament as it is extruded, we aim to induce a radiative force capable of counteracting the force of gravity as well as spatially stabilizing the filament in an acoustic potential. The overall vision is to create a sustainable method to support complex shapes at various angles during printing.

We analyze and demonstrate a device where contactless acoustic support is integrated into linear 3D printing. Our approach mimics the effect of solid supports in 3D printing by optimally distributing the acoustic radiation force across the filament as it hardens (**FIGURE 1**). The acoustic radiation force (ARF) [1-4] arises from the momentum change of sound as it scatters from an object and has been used in concepts such as acoustic tweezers and traps [5-14]. Such traps have been used to levitate small animals [15], fluid droplets [16-19], in chemical analysis [20] and medical devices [21]. These acoustic levitators establish symmetrical resonant acoustic traps to levitate single, spherical particles. It has been shown that these devices can be used to assemble larger shapes using individual particles as building blocks [22]. Rather than analyzing the concatenation of such particles, here we explore a method for the support of continuous solid manufacturing during filament extrusion. In doing so, it could be possible to achieve finer surface detail and functionally stronger objects. We expect the continued study of contactless additive manufacturing to introduce new alternatives to solid support structures in multiple dimensions, with the potential to replace the standard printer bed in some applications.

We explore the applications of acoustic levitation devices in contactless fabrication as they demonstrate functionality under a wide variety of materials [23-24] and offer means to be uniquely designed based on the levitated object. Our paper is outlined as follows. First, we analyze the stability of a typical 3D-printer filament in a controllable ultrasonic acoustic field. Next, we design a full acoustic levitator by evaluating complex acoustic potentials using perturbation theory and finite-element simulations. We then confirm the numerical design and assemble a functional device. Finally, we pair the acoustic levitation device with conventional fused filament fabrication (FFF) 3D printing to show that the linear trapping of solid and semisolid objects is possible during extrusion.

**METHODS**

We identify recurring characteristics across existing methods for acoustic levitation. These devices often employ a configuration where the sound source is reflected to create a standing wave [25-26], with nodal traps that support the condition for levitation. We further tailor the nodal traps by arranging an array of transducers in a concave shape [27]. Focusing the emitted sound waves from either side of the levitation device allows for the increased distance between each emitter without compromising the maximum radiation force applied. Concave transducer arrays have been used in previous devices for levitation of spherical objects; however, these devices often form spherical caps [28-29], limiting the shape and size of trapped objects. To overcome this limitation, we design a non-resonant cylindrical array of transducers capable of levitating linear objects up to the length of the array.

To design our array, we judiciously select the arrangement of transducers with the goal of maximizing output pressure and minimizing diaphragm radius. The operating frequency of 40 kHz is selected as this frequency is inaudible to the human ear and gives a maximum trap size of ≈ 4.3 mm ($\lambda/2$). We chose this frequency based on the knowledge that it generates sufficiently sized acoustic traps for our purposes. Common FDM 3D printers can be equipped with nozzle diameters ranging from 0.1 mm to 2.0 mm. This indicates that the filament will fit within a single acoustic trap while maintaining enough leeway to permit slight offsets in the alignment of the array and extrusion nozzle. Both transducer arrays are wired in parallel and driven by an Arduino Nano emitting a square wave in conjunction with a driver amplifier. A Creality Ender 3 3D printer was modified to extrude filament horizontally inside the levitation device, minimizing downward force applied by the extruder. This printer was chosen as a low-cost option that can easily interface with Pronterface, a free GUI allowing full real-time control of each stepper motor.

We utilize COMSOL Multiphysics to establish finite element simulations, allowing us to predict the behavior and performance of our array [30]. As discussed in this work, we numerically compute the ARF components using the momentum flux evaluated over a closed contour integral existing outside the surface of the levitated object. These simulations were performed in 2D for two separate planes of interest (2D simulations were chosen to reduce computation time while maintaining confidence in the synthesized array). The first plane analyzed is the circular cross-section of the cylindrical array, which we label as the x-y plane; the second plane exists along the length of the array, labeled the y-z plane. A circular hard boundary representative of PLA filament (radius of 0.2 mm) was positioned throughout the array, and the radiation force was evaluated. From the x-y plane, we establish the trap location, radiation force magnitude, and array diameter. The y-z plane allows us to evaluate optimal transducer spacing along the length of the cylindrical array. Combining the results from each plane fully defines the geometric constraints of the proposed levitation array.

**RESULTS**

We design a cylindrical non-resonant array for use in linear contactless support. The choice of array shape was selected to establish stable levitation of the filament from all sides along the length of extrusion. The proposed array was evaluated for maximum acoustic radiation force while maintaining an evenly distributed force over the surface of the filament. The cross-section profile of the array will determine the magnitude of applied radiation force. We first select a cylindrical radius of 45 mm based on the output frequency and diameter of the selected transducers. This radius ensures that standing waves will be present within the array. We note that increasing the radius of the cylindrical array allows more transducers to be placed along the circumference of the array but reduces the applied radiative force from each individual transducer due to an increasing distance between the levitated object and the sources of pressure. By selecting a radius of 45 mm, we maintain high acoustic radiation force while minimizing the total number of necessary transducers. To simplify the integration of the acoustic array and a 3D printer, we remove several transducers from the circumference of the array, therefore creating a gap (40 mm) that allows a nozzle to reach into the intended acoustic trap. **FIGURE 2a** shows the cross-sectional profile of the array and the corresponding acoustic field profile obtained in COMSOL.

Defining the cross-sectional geometry of the array allows us to determine ranges for filament size and output pressure in which stable support can be achieved. Filament radius and harmonic output pressure were varied, and the resulting acoustic radiation force was recorded. We then examine how the radiation force relates to the force of gravity on polylactic acid filament (density of 1.25 g/cm$^3$), a commonly used material in FFF printing. **FIGURE 2b** shows the numerically established relationship between the radiation force and the gravitational force on the filament. We observe that low output pressures are not able to provide sufficient contactless support (i.e., the force of gravity is greater than the acoustic force); however, as the output pressure increases, a regime of potential trapping is realized (i.e., the acoustic force is greater than the force of gravity). We indicate that a harmonic output pressure of 250 Pa per transducer provides a reasonable estimate of the selected transducers while ensuring a levitation range over the desired nozzle sizes. In the subsequent analysis of trap regions and transducer spacing, we select a filament radius of 0.2 mm which is a common nozzle size used in 3D printing.

We determine the optimal locations for filament trapping and the trap size by numerically analyzing the components of the acoustic force on the filament. To do this, we sweep the position of the filament on the X-Y cross-section plane and record the lateral trapping force ($F_x$) and the vertical supporting force $F_y$ that is antiparallel to gravity (note that Z-axis is the extrusion axis). **FIGURE 3a** shows the profile of the vertical force. We expect that a suitable trapping location would be located along the centerline of the trap at a vertical height between $\frac{\lambda}{8}$ and $\frac{\lambda}{4}$ from the center of the array. We observe the maximum value of $F_y$ is achieved along the $x = 0$ line, which indicates that the filament would be supported near these coordinates. To assess whether there is sufficient lateral/horizontal force, we refer to **FIGURE 3b,** which depicts $F_x$ as a function of the filament position. Indeed, we observe that the profile of the lateral force is stabilizing – that is, $F_x > 0$ for $x < 0$ and $F_x < 0$ for $x > 0$. By evaluating the radiation force exerted on filament at different positions throughout the array, we also observe the overall size of the acoustic trap, which extends approximately 2 mm in either direction horizontally and 1 mm in either direction vertically. A greater trap size is key to the success of supporting the filament during extrusion as it allows for leeway in the nozzle placement. Specifically, this enables for off-center extrusion to be corrected and fully supported without recalibration of the printing device.

Besides having a sufficient trapping region for the nozzle, a successful continuous extrusion of filament requires evenly distributed acoustic radiation force, which is dictated by the spacing between circular rows within the array. By evaluating several transducer spacings, we assess an optimal range of locations in which transducers can be placed to create an effective linear trap. Transducer spacing ranging from 0 mm to 5 mm were investigated, and the force trends are shown in **FIGURE 4a**. We observe that the acoustic force remains positive and relatively uniform for spacings up to 4 mm. Beyond this region, nodal acoustic traps begin to form along the z-axis. Low-pressure regions exist between these nodes, causing the filament to experience a downward acoustic force as seen in **FIGURE 4c**. These nodal traps promote the levitation of spherical subwavelength objects but are detrimental in a linear acoustic trap as they do not provide a positive force along the entire length of the filament. A final spacing of 3 mm was chosen for the synthesized array, as seen in **FIGURE 4b**. This spacing provides a relatively uniform force distribution while increasing the length of the array and, in turn, allowing longer filament lengths to be printed for a given number of transducers. We comment that the forces observed in **FIGURE 4** are primarily used to display the trends that develop as transducer spacing increases; the actual magnitude of the force is not of consequence for analyzing the trends.

Having established a suitable geometry and arrangement of acoustic sources, we proceed to build a concept prototype and test it under a variety of filament diameters and printing conditions (Supplementary Materials). To test the predictions of our numerical finite element simulations, a single strand of pre-extruded PLA filament with a radius of 0.2 mm was first successfully levitated in the array. The results of this test can be found in the Supplementary Material. We then proceed to examine the effects of contactless acoustic support on 3D printing extrusion. In our tests, the extrusion is initiated from a vertical base positioned on the beginning end of the array. **FIGURE 5a** shows a line of filament being extruded over time when the transducer array is turned off as a control case. **FIGURE 5b** shows extrusion when the transducer array is turned on. Full videos of each print can be found in the Supplementary Materials. We observe a clear advantage of supporting the filament during extrusion: without the air bed support, the filament sags down to a depth of 14 mm; in contrast, when the air bed is present, there is a much smaller (3 mm) deviation in the vertical position of the filament. Supplementary Materials contain additional figures of the filament during extrusion. We remark that this is intended as a prototype concept demonstration: further improvement in acoustic field profile and printing conditions would help improve the quality and stability of extrusion. It should also be noted that, for simplicity, our simulations do not account for the presence of the nozzle and the extrusion head, which could lead to discrepancies between the theory and the experiment. Additionally, it was assumed that all transducers have the same phase and pressure output when driven by an identical signal, an assumption that may not always hold in a real system and could affect the performance of the air bed support during extrusion.

## SUMMARY

In summary, we analyzed a method for the contactless support of linear additive manufacturing via an acoustic air bed. By introducing a tailored acoustic potential, we show that a commonly-used 3D material filament can be supported during extrusion in a prototype device that integrates a commercial 3D printer with open-source control code. Optimal system geometry and acoustic source arrangements are studied through finite-element acoustic simulations. This device expands upon previous works in acoustic trapping by investigating the conditions for and realizing uniformly distributed linear acoustic potentials suitable for material extrusion and analyzing filament dynamics in the acoustic field. The concept of 3D printing on an air bed could lead to fewer surface defects, reduced material waste, cost, and manufacturing time associated with the generation of solid supports in conventional 3D printing.

## References

[1] Bruus, Henrik. "Acoustofluidics 7: The Acoustic Radiation Force on Small Particles." *Lab on a Chip*, vol. 12, no. 6, 21 Feb. 2012, p. 1014.

[2] Settnes, Mikkel, and Henrik Bruus. "Forces Acting on a Small Particle in an Acoustical Field in a Viscous Fluid." *Physical Review E*, vol. 85, no. 1, 2012.

[3] Andrade, Marco A., et al. "Acoustic Levitation of a Large Solid Sphere." *Applied Physics Letters*, vol. 109, no. 4, 2016, p. 044101., https://doi.org/10.1063/1.4959862.

[4] Glynne-Jones, Peter, et al. "Efficient Finite Element Modeling of Radiation Forces on Elastic Particles of Arbitrary Size and Geometry." *The Journal of the Acoustical Society of America*, vol. 133, no. 4, 2013, pp. 1885–1893.

[5] Shi, Qianqian, et al. "A General Approach to Free-Standing Nanoassemblies *via* Acoustic Levitation Self-Assembly." *ACS Nano*, vol. 13, no. 5, 2019, pp. 5243–5250., https://doi.org/10.1021/acsnano.8b09628.

[6] Marzo, Asier, et al. "TinyLev: A Multi-Emitter Single-Axis Acoustic Levitator." *Review of Scientific Instruments*, vol. 88, no. 8, 2017, p. 085105.

[7] Marzo, Asier, et al. "Ultraino: An Open Phased-Array System for Narrowband Airborne Ultrasound Transmission." *IEEE Transactions on Ultrasonics, Ferroelectrics, and Frequency Control*, vol. 65, no. 1, 2018, pp. 102–111.

[8] Marzo, Asier, et al. "Holographic Acoustic Elements for Manipulation of Levitated Objects." *Nature Communications*, vol. 6, no. 1, 2015.

[9] Vandaele, Vincent, et al. "Non-Contact Handling in Microassembly: Acoustical Levitation." *Precision Engineering*, vol. 29, no. 4, 2005, pp. 491–505., https://doi.org/10.1016/j.precisioneng.2005.03.003.

[10] Wu, Huaying, et al. "Design of Ultrasonic Standing Wave Levitation Support for Three-Dimensional Printed Filaments." *The Journal of the Acoustical Society of America*, vol. 149, no. 4, 28 Apr. 2021, pp. 2848–2853., https://doi.org/10.1121/10.0003922.

[11] Trinh, E. H. "Compact Acoustic Levitation Device for Studies in Fluid Dynamics and Material Science in the Laboratory and Microgravity." *Review of Scientific Instruments*, vol. 56, no. 11, 9 Sept. 1998, pp. 2059–2065., https://doi.org/10.1063/1.1138419.

[12] Andrade, Marco A., et al. "Acoustic Levitation of an Object Larger than the Acoustic Wavelength." *The Journal of the Acoustical Society of America*, vol. 141, no. 6, 6 June 2017, pp. 4148–4154., https://doi.org/10.1121/1.4984286.

[13] Kandemir, Mehmet Hakan, and Mehmet Çalışkan. "Standing Wave Acoustic Levitation on an Annular Plate." *Journal of Sound and Vibration*, vol. 382, 2016, pp. 227–237., https://doi.org/10.1016/j.jsv.2016.06.043.

[14] Zhao, Su, and Jörg Wallaschek. "A Standing Wave Acoustic Levitation System for Large Planar Objects." *Archive of Applied Mechanics*, vol. 81, no. 2, 2009, pp. 123–139., https://doi.org/10.1007/s00419-009-0401-3.

[15] Xie, W. J., et al. "Acoustic Method for Levitation of Small Living Animals." *Applied Physics Letters*, vol. 89, no. 21, 2006, p. 214102.

[16] Priego-Capote, F., and Luque De Castro. "Ultrasound-Assisted Levitation: Lab-on-a-Drop." *TrAC Trends in Analytical Chemistry*, vol. 25, no. 9, 2006, pp. 856–867., https://doi.org/10.1016/j.trac.2006.05.014.

[17] Watanabe, Ayumu, et al. "Contactless Fluid Manipulation in Air: Droplet Coalescence and Active Mixing by Acoustic Levitation." *Scientific Reports*, vol. 8, no. 1, 2018, https://doi.org/10.1038/s41598-018-28451-5.

[18] Foresti, Daniele, et al. "Acoustophoretic Contactless Transport and Handling of Matter in Air." *Proceedings of the National Academy of Sciences*, vol. 110, no. 31, 2013, pp. 12549–12554., https://doi.org/10.1073/pnas.1301860110.

[19] Nikolaeva, A.V., et al. "Acoustic Radiation Force of a Quasi-Gaussian Beam on an Elastic Sphere in a Fluid." *2016 IEEE International Ultrasonics Symposium (IUS)*, 2016.

[20] Santesson, Sabina, and Staffan Nilsson. "Airborne Chemistry: Acoustic Levitation in Chemical Analysis." *Analytical and Bioanalytical Chemistry*, vol. 378, no. 7, 2004, pp. 1704–1709., https://doi.org/10.1007/s00216-003-2403-2.

[21] Xie, W. J., and B. Wei. "Parametric Study of Single-Axis Acoustic Levitation." *Applied Physics Letters*, vol. 79, no. 6, 2001, pp. 881–883., https://doi.org/10.1063/1.1391398.

[22] Ezcurdia, Iñigo, et al. "LeviPrint: Contactless Fabrication Using Full Acoustic Trapping of Elongated Parts." *Special Interest Group on Computer Graphics and Interactive Techniques Conference Proceedings*, 2022.

[23] Andrade, Marco A., et al. "Review of Progress in Acoustic Levitation." *Brazilian Journal of Physics*, vol. 48, no. 2, 2017, pp. 190–213., https://doi.org/10.1007/s13538-017-0552-6.

[24] Foresti, Daniele, et al. "Investigation of a Line-Focused Acoustic Levitation for Contactless Transport of Particles." *Journal of Applied Physics*, vol. 109, no. 9, 2011, p. 093503., https://doi.org/10.1063/1.3571996.

[25] Xie, W. J., and B. Wei. "Temperature Dependence of Single-Axis Acoustic Levitation." *Journal of Applied Physics*, vol. 93, no. 5, 2003, pp. 3016–3021., https://doi.org/10.1063/1.1540232.

[26] Andrade, Marco A., et al. "Nonlinear Characterization of a Single-Axis Acoustic Levitator." *Review of Scientific Instruments*, vol. 85, no. 4, 2014, p. 045125., https://doi.org/10.1063/1.4872356.

[27] Morris, Robert H., et al. "Beyond the Langevin Horn: Transducer Arrays for the Acoustic Levitation of Liquid Drops." *Physics of Fluids*, vol. 31, no. 10, 2019, p. 101301., https://doi.org/10.1063/1.5117335.

[28] Xie, W. J., and B. Wei. "Dependence of Acoustic Levitation Capabilities on Geometric Parameters." *Physical Review E*, vol. 66, no. 2, 2002, https://doi.org/10.1103/physreve.66.026605.

[29] Karlsen, Jonas T., and Henrik Bruus. "Forces Acting on a Small Particle in an Acoustical Field in a Thermoviscous Fluid." *Physical Review E*, vol. 92, no. 4, 2015.

[30] Stein, Keller, Luo, and Ilic "Shaping contactless force through anomalous acoustic scattering", 2022, arXiv.2204.04137

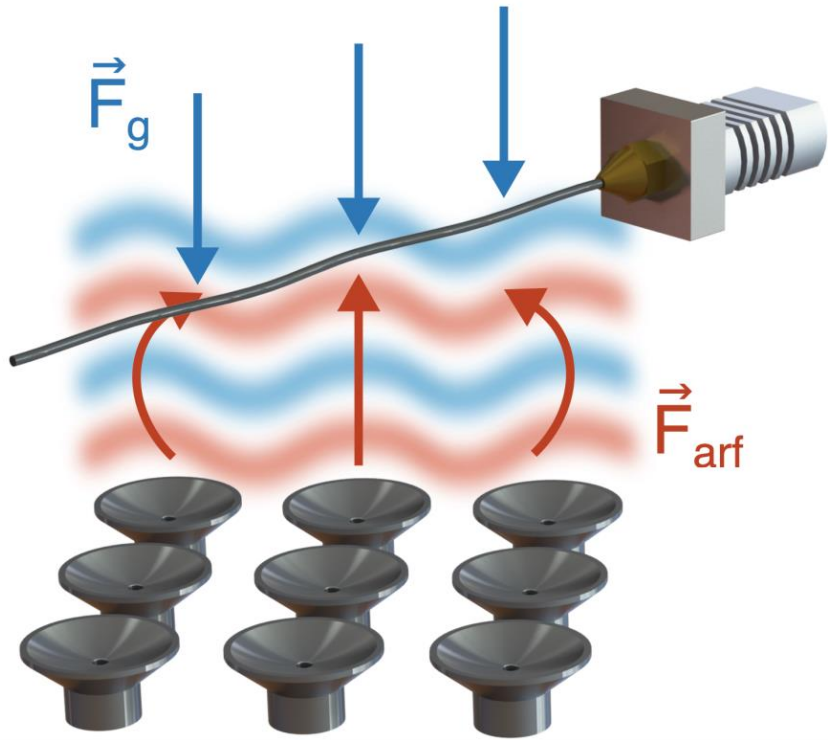

**FIGURE 1. Contactless 3D printing via material extrusion on an acoustic air bed.** Concept schematic of directing a tailored acoustic wavefront to contactlessly support material during extrusion. Engineered arrangement of transducers induces an acoustic radiative force capable of counteracting the force of gravity as well as spatially stabilizing the extruded material in an acoustic potential.

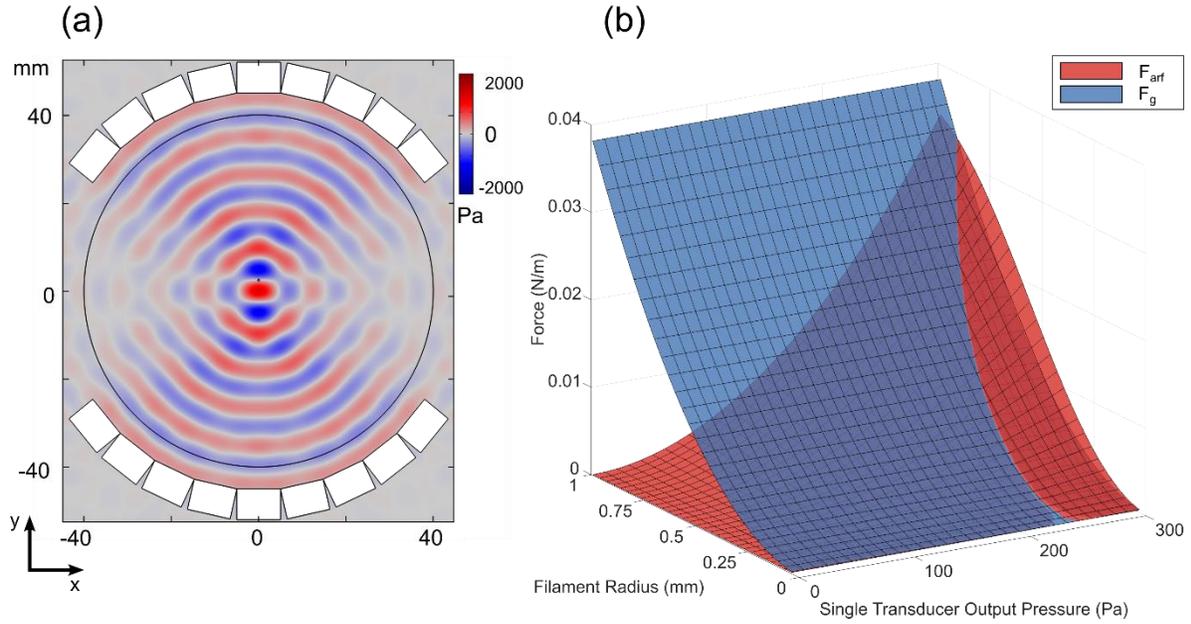

**FIGURE 2. Numerical finite-element acoustic model and analysis of the acoustic radiation force. (a)** Cross-section of the transducer array and the resulting acoustic field profile (contour-plot). Y-axis denotes gravity (pointing along -Y), Z-axis is the extrusion direction (out of plane). Black circle denotes the integration boundary for evaluating the radiation force on the filament. **(b)** Vertical acoustic force ($F_y$) and the gravitational force ($F_g$) as a function of the filament radius and the output transducer pressure. A functional region in which applied radiative force is greater than the force of gravity is established for the specified array geometry in (a).

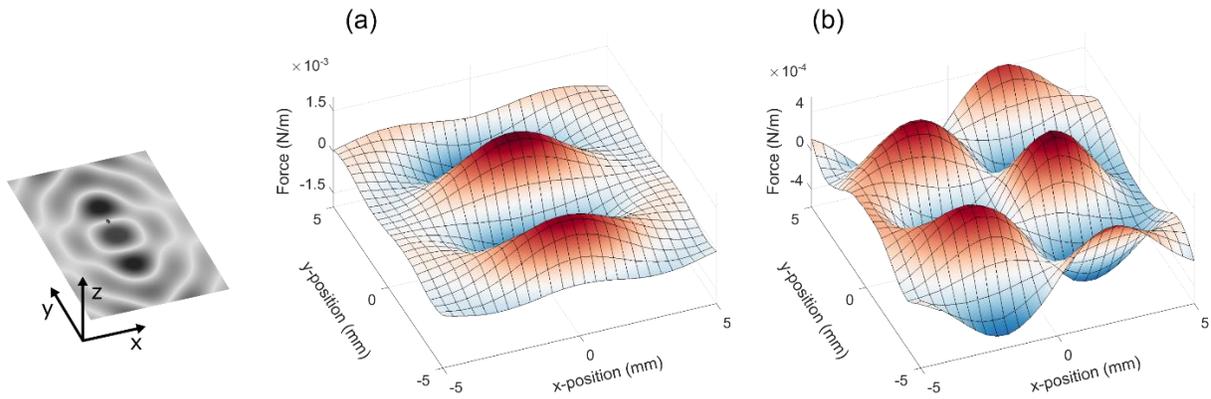

**FIGURE 3. (a) Vertical radiation force exerted on the circular filament cross section. (b) Horizontal force exerted on the filament.** The magnitude of applied acoustic radiation force is shown along with the exact filament trapping location. Vertical force is significantly greater than horizontal force to ensure stable support in standard gravity.

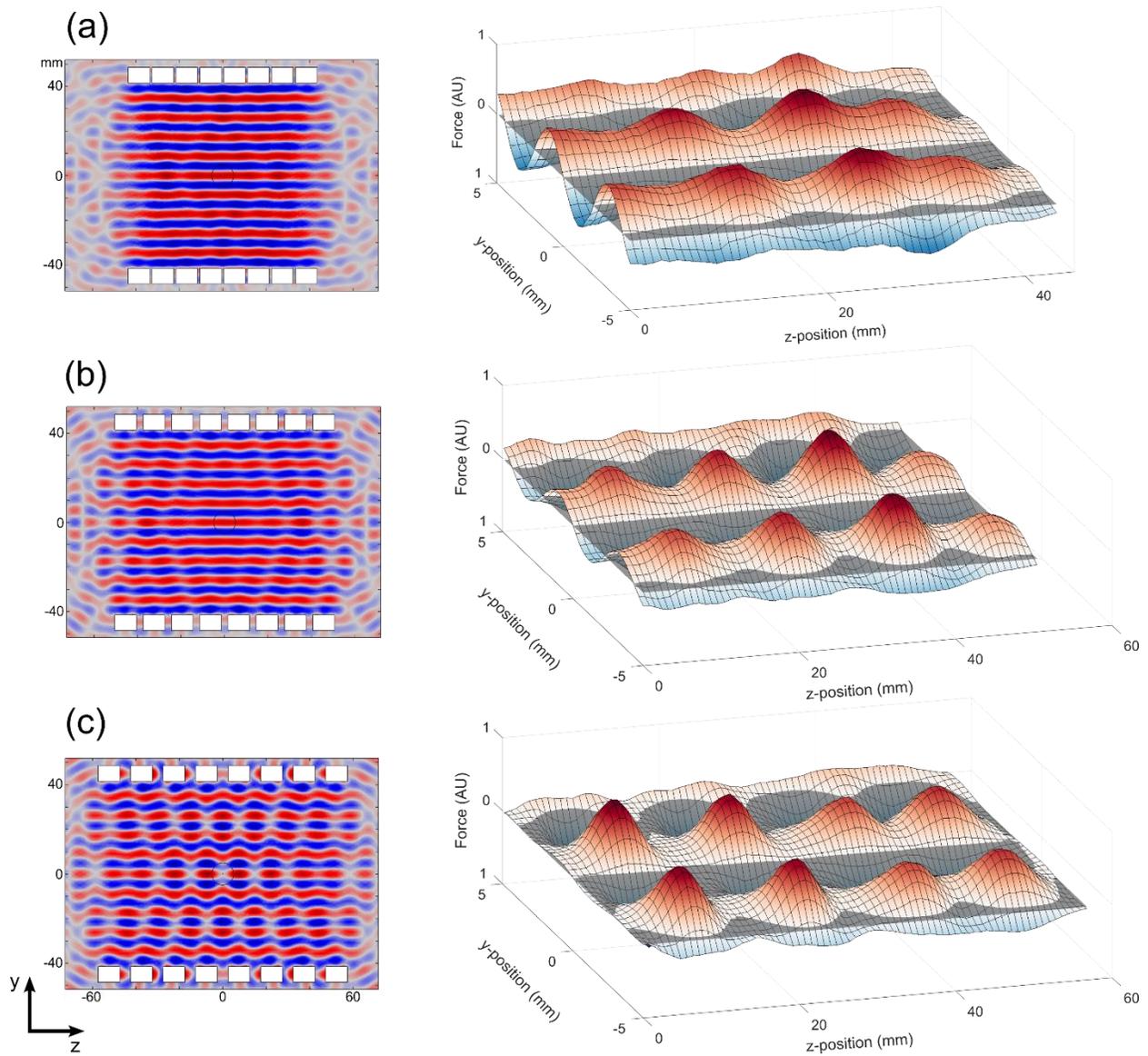

**FIGURE 4. Analysis of optimal transducer array spacing.** Vertical acoustic radiation force ($F_y$) as a function of position on the plane comprising the extrusion direction (Z) and the gravity direction (-Y). **(a)** Transducer spacing is 1 mm; **(b)** Transducer spacing is 3 mm; **(c)** Transducer spacing is 5 mm. When transducer spacing is increased, the vertical acoustic force $F_y$ becomes weaker and less uniform along the extrusion direction (Z), even turning negative in some locations along the filament. The displayed trend points to an optimal tradeoff between force uniformity/magnitude and transducer density.

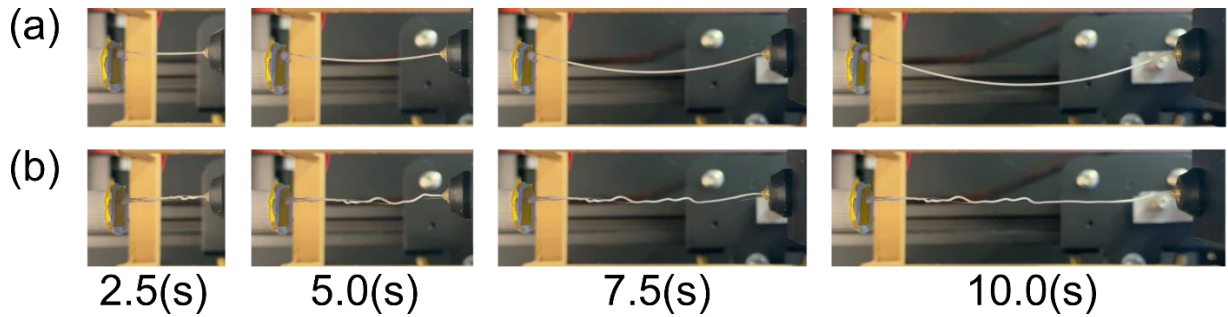

**FIGURE 5. Experimental dynamics of extrusion of a 3D-printer filament. (a)** Acoustic field is turned OFF, and the filament sags under its weight as it is extruded. **(b)** Acoustic field is turned ON; the filament is supported by an air bed and better maintains its shape. The acoustic radiation force improves the shape of printing without solid supports. The horizontal axis denotes snapshots at different times.